\def\fr#1#2{\hbox{${#1\over #2}$}}  
\def\subsub#1{\paragraph*{#1}}
\def\Tr{\,{\rm Tr}\,}    
\def\pd{\partial}        
\def\m{\mu}              \def\n{\nu}              \def\k{\kappa}
\def\G{\Gamma}           \def\S{\Sigma}           \def\s{\sigma} 
\def\a{\alpha}           \def\vphi{\varphi}       
\def\cs{\omega}                     
\def\O{\Omega}           \def\bO{\bar\Omega}
\def\bv{{\bar v}}        \def\bg{{\bar g}{}}
            
\def\gp{ {(+)} }         \def\gn{ {(0)} }         \def\gm{ {(-)} }     
\documentstyle[12pt]{article}

\def\nn{\nonumber}                    \def\ed{\end{document}}
\def\be{\begin{equation}}             \def\ee{\end{equation}}
\def\bea{\begin{eqnarray} }           \def\eea{\end{eqnarray} }
\def\lab#1{\label{#1}}                \def\eq#1{(\ref{#1})}
\def\bitem{\begin{itemize}}           \def\eitem{\end{itemize}}  
\begin{document}
\begin{center}
{\Large\bf Gauge connection between the WZNW system \\
           and 2D induced gravity\footnote{A talk presented at the
XI International Conference {\it Problems of Quantum Field Theory}, in 
memory of D. I. Blokhintsev, Dubna, July 13--17, 1998. }}

\bigskip
{\large M. Blagojevi\'c\footnote{ 
                           E--mail address: mb@phy.bg.ac.yu}
             and \underline{B. Sazdovi\'c}\footnote{ 
                           E--mail address: sazdovic@phy.bg.ac.yu} } 
\date{}

\smallskip
{\it Institute of Physics, 11001 Belgrade, P.O.Box 57, Yugoslavia}       
\end{center}

\begin{abstract} 
We introduce a consistent gauge extension of the $SL(2,R)$ WZNW system,
defined by a difference of two simple WZNW actions. By integrating out
some dynamical variables in the functional integral, we show that the
resulting effective theory coincides with the induced gravity in 2D. 
General solutions of both theories are found and related to each other.
\end{abstract}
\section*{\large\bf 1~~Gauge extension of the WZNW system} 

Dynamical structure of two--dimensional (2D) gravity is an important
aspect of string theory \cite{1}, but it also represents a useful model
for the theory of gravitational phenomena in four dimensions. An
interesting connection between 2D induced gra\-vi\-ty and the $SL(2,R)$
Wess--Zumino--Novikov--Witten (WZNW) model has been discovered both in
the {\it light--cone\/} gauge \cite{2} and in the {\it conformal\/}
gauge \cite{3,4}. Here, we present our recent result on the subject
\cite{5}, showing that the related connection can be established in a
{\it covariant\/} way, fully respecting the diffeomorphism invariance
of the induced gravity. We formulate a consistent gauge theory of the
WZNW system,  
\be
S(g_1,g_2)=S(g_1)-S(g_2)\, ,\qquad g_1,g_2\in SL(2,R)\, ,      \lab{1}
\ee
defined by a difference of two simple WZNW actions, and show, by
integrating out some dynamical variables in the functional integral,
that the resulting effective theory reduces to the induced gravity
in 2D: 
\be
S_G(\phi,g_{\m\n})=\int d^2\xi \sqrt{-g}\,\bigl[ 
    \fr{1}{2}g^{\m\n}\pd_\m\phi\pd_\n\phi
   +\fr{1}{2}\a\phi R -M\bigl( e^{2\phi/\a} -1\bigr) \bigr]\, .\lab{2}
\ee
General solutions of both theories  are constructed, and the related
connection between them is discussed \cite{6}.

\subsub{WZNW model.} WZNW model in 2D is a field theory described by
the action 
\be
S(g)=S_0(v) + n\G(v) 
      =\fr{1}{2}\k\int_\S ({}^*v,v) + \fr{1}{3}\k\int_M (v,v^2) 
      \qquad          (\k= n\k_0)\, .                          \lab{3}
\ee
The first term is the action of the non--linear $\s$--model for a
group valued field $g$: $\S \to G$, where $\S$ is two--dimensional
Riemannian manifold (spacetime) and $G$ a semi--simple Lie group.  
Here, $v=g^{-1}dg$ is the Maurer--Cartan (Lie algebra valued)
1--form, ${}^*v$ is the dual of $v$, and $(X,Y)=\fr{1}{2}\Tr(XY)$ is
the Cartan--Killing bilinear form.  
The second term is the topological Wess--Zumino term, defined on a 
three--manifold $M$ whose boundary is $\S=\pd M$. With a suitable
choice of the normalization constant $\k_0$ the Wess--Zumino term is
well defined modulo a multiple of $2\pi$.

The action $S(g)$ is invariant under the {\it global\/} transformations
$g\to g'=\O g\bO^{-1}$, where $(\O,\bO)$ belongs to $G\times G$. 
We want to introduce the corresponding gauge theory, which is 
\bitem
\item[(a)] invariant under {\it local\/} transformations, and
\item[(b)] defined as a field theory over spacetime $\S$.   
\eitem
To do that, we introduce the {\it covariant derivative\/}: 
\be
Dg\equiv dg+Ag-gB \, ,\qquad (Dg)'=\O(Dg)\bO^{-1} \, ,         \lab{4}
\ee
where $A$ and $B$ are {\it gauge fields\/} (Lie algebra valued
1--forms). The covariant derivative $Dg$ transforms homogeneously under
local transformations, provided the gauge fields transform according to   
$A'=\O (A + d)\O^{-1}$, $B'=\bO (B + d)\bO^{-1}$.

Now, by replacing 1--form $v=g^{-1}dg$ with the corresponding
covariant 1--form,  
$$
v\to V\equiv g^{-1}Dg=v +g^{-1}Ag -B\, ,\qquad V'=\bO V\bO^{-1}\, , 
$$
we can formally define a gauge invariant extension of the WZNW action
\eq{3}:   
\be
S(g,A,B)=S_0(V)+n\G(V)
    =\fr{1}{2}\k\int_\S ({}^*V,V)+\fr{1}{3}\k\int_M (V,V^2)\, . \lab{5}
\ee
Although the action \eq{5} is gauge invariant, it is not defined as
a field theory over two--dimensional manifold $\S$. The reason for this
lies in the form of the second term: 
\bea
&&n\G(V)=n\G(v)+\G_1+\G_2+\G_3 \, ,\nn \\
&&\G_1=\k \int_\S \fr{1}{2}\Tr\bigl[
       -\bv A + vB +g^{-1}AgB \bigr] \, , \nn \\
&&\G_2=\k \int_M\fr{1}{2}\bigl[\cs_3(B,F_B)-\cs_3(A,F_A)\bigr]\, ,\nn \\
&&\G_3=\k \int_M \fr{1}{2}\Tr\bigl[
      F_{A}(Dg)g^{-1} +F_{B}g^{-1}(Dg) \bigr] \, ,      \nn 
\eea
where $\cs_3(A,F_A)=\Tr(AF_{A}-\fr{1}{3}A^3)$ is the Chern--Simons
3--form. We see that $\G_2$ and $\G_3$ are defined not on $\S$ but
on $M$, violating thereby the basic requirement (b). Therefore, the
expression \eq{5} is not an acceptable gauge extension of $S(g)$.  

\subsub{WZNW system.} The following solution of this problem has been
given in Ref. \cite{5}. One first observes that the term $\G_3$ is
gauge invariant, therefore it can be removed from $S(g,A,B)$, leaving
us with the gauge invariant expression $S'=S^r+\G_2$, where $S^r$ is
the regular part of $S$, which is (a) gauge invariant and (b) defined
on $\S$. Now, using the fact that $\G_2$ does not depend on $g$, one
can solve the $\G_2$ problem by going over to the {\it WZNW system\/},
given by a difference of two simple WZNW models, where the
contributions of two $\G_2$ terms mutually cancel. Thus, a consistent
gauge invariant extension of the WZNW system is given by    
\be
S(1,2)=S^r(g_1,A,B)-S^r(g_2,A,B) \, ,                          \lab{6}
\ee
where $g_1$ and $g_2$ are two independent fields, and the regular part
has the form  
\be
S^r(g,A,B) =S(v) +\k \int_\S d^2\xi\sqrt{-g}\Tr\bigl[ 
            -\bv_-A_+ -v_+B_- -B_-g^{-1}A_+g \bigr]\,  .       \lab{7}
\ee
Here, $\bv=gdg^{-1}$, $g_{\m\n}$ is the metric tensor on $\S$, and
$U_{\pm}$ are vector components in the light-cone basis \cite{5}.   

The action \eq{6} does not contain variables $(A_-,B_+)$, which implies
the existence of an extra gauge symmetry. To simplify further
considerations we shall fix this symmetry by demanding $A_-=B_+=0$.

The independent fields in the action \eq{6} are: $g_1$, $g_2$, $A_+$,
$B_-$, and only two components of the metric tensor (because of the
invariance of $S^r$ under conformal rescalings), which we
conveniently choose as $h^\pm=(-g_{01}\pm\sqrt{-g})/g_{11}$ .

\section*{\large\bf 2~~Transition to the induced gravity} 

We shall now give an explicit formulation of the gauged $SL(2,R)$ WZNW
system \eq{1}, and see that this theory, after a suitable gauge fixing
and integrating out some dynamical variables, leads to the induced
gravity action \eq{2}.   

\subsub{$H_+\times H_-$ gauge theory.} Let us focus our
attention to the case $G=SL(2,R)$, and choose the generators of $SL(2,R)$ as  
$t_{(\pm)}=\fr{1}{2}(\s_1\pm i\s_2)$ and  $t_\gn =\fr{1}{2}\s_3$,  where
$\s_k$ are the Pauli matrices. Next, we observe that any element $g$ of
$SL(2,R)$ in a neighborhood of identity admits the Gauss decomposition, 
\be
g = e^{x t_{(+)}} e^{\vphi t_{(0)}} e^{y t_{(-)}} 
  = e^{-\vphi/2}  \pmatrix{ e^{\vphi}+xy   &  x  \cr
                                  y        &  1  \cr }\, ,    \lab{8}
\ee
where $(x,\vphi,y)$ are group coordinates. This completely defines
the action \eq{1} of the WZNW system for $SL(2,R)$.

Although we could take the gauge group $H$ to be the whole 
$SL(2,R)\times SL(2,R)$, this is not necessary. We assume that $H$ is
given as $H=H_+\times H_-$, where $H_+$ and $H_-$ are subgroups of
$SL(2,R)$ defined by the generators $(t_+,t_0)$ and $(t_0,t_-)$,
respectively. When compared to $SL(2,R)\times SL(2,R)$, our choice
means that the gauge fields should be restricted as follows:
\be
A^\gm_+=0 \,  , \qquad  B^\gp_- =0 \,   .                      \lab{9}
\ee
The action of the gauged $SL(2,R)$ WZNW system, defined in terms of 
$(x_1,\vphi_1,y_1)$, $(x_2,\vphi_2,y_2)$,
$(A^\gp_+,A^\gn_+,B^\gn_-,B^\gm_-)$ and $(h^+,h^-)$, has the form 
\eq{6}, where 
\be
S^r(g,A,B) = \k\int d^2\xi\sqrt{-g}\,\bigl[ \pd_-\vphi\pd_+\vphi
 +2A^\gn_{+}\pd_-\vphi -2B^\gn_{-}\pd_+\vphi 
 +4D_+xD_-y\,e^{-\vphi} \bigr] \, ,                           \lab{10}
\ee
and $D_+ x =\bigl[\pd_+ +A^\gn_+\bigr]x +A^\gp_+$, 
    $D_- y =\bigl[\pd_- -B^\gn_-\bigr]y -B^\gm_-$,
are covariant derivatives on the group manifold. 
\subsub{Effective theory.} In order to demonstrate the connection of
the gauged WZNW system to the induced gravity, we shall rewrite the
part of the action \eq{6}, given by 
$D_+x_1 D_- y_1 e^{-\vphi_1}-(1\to 2)$, in the form
$$
f_{1-}D_+x_1 +f_{1+}D_-y_1 -\fr{1}{4}f_{1-}f_{1+}e^{-\vphi_1} 
                                                    - (1\to 2) \, ,
$$
where $f_{1\pm},f_{2\pm}$ are auxiliary fields. Now, the construction
of the effective theory goes as follows: 
\bitem
\item[$i)$] first, we impose the partial gauge fixing $x_2=0$, $y_2=0$,
and integrate over $(A^\gp_+,B^\gm_-)$, $(f_{2+}f_{2-})$, $(x_1,y_1)$
and $(A^\gn_+,B^\gn_-)$;  
\item[$ii)$] then, we introduce new variables $\phi_1$ and $\phi_2$ by
\be
\m e^{\phi_1}= f_{1+}f_{1-} e^{\vphi_1}\, ,\qquad 
\m e^{\phi_2}= f_{1+}f_{1-} e^{\vphi_2}\, ,                  \lab{11}
\ee
($\m$ is a normalization constant), which are invariant under conformal
rescalings.   
\eitem
The geometric meaning of the effective theory obtained in this way
becomes more transparent if we change the variables according to  
\be
\phi=\sqrt{\k}\bigl(\phi_1-\phi_2\bigr) \, , 
           \qquad \bg_{\m\n}=g_{\m\n} e^{\phi_2} \, .        \lab{12}
\ee
The effective Lagrangian is seen to coincide with the induced
gravity Lagrangian \eq{2}, where $\a=2\sqrt{\k}$, $M=\k\m$, and
$\bg_{\m\n}$ plays the role of the metric tensor of the induced
gravity.  The metric tensor $\bg_{\m\n}$ now has three independent
components. 

\section*{\large\bf 3~~Conclusions}

We presented here the connection between the gauge extension of the
WZNW system \eq{1} and the induced gravity action \eq{2}, fully
respecting the diffeomorphism invariance of the induced gravity.
Our gauge group is $H_+\times H_-$, a four--parameter subgroup of
$SL(2,R)\times SL(2,R)$. In the process of constructing the induced
gravity action, we found a natural explanation of the geometry of
spacetime in terms of the gauge structure of the WZNW system.   

These results can be used to improve our understanding of singular
solutions of 2D induced gravity in terms of globally regular solutions
of the WZNW system \cite{4}. The equations of motion obtained from the
$SL(2,R)$ WZNW action, defined by equations \eq{6} and \eq{10}, have
the following general solution \cite{6}:   
\be
g_1 = G_- G_+  \,  , \qquad   g_2 = G_- e G_+  \,  .         \lab{13}
\ee
Here, $G_{\pm}$ are functions on $\S$ belonging to $SL(2,R)$, such that 
\be
D_{\pm} G_{\mp} =0 \, ,                                      \lab{14}
\ee
while $e$ is a constant matrix in $SL(2,R)$.
For arbitrary gauge fields $A_+$, $B_-$, and any metric tensor
$g_{\m\n}$ on $\S$, one can solve equations \eq{14} for $G_\mp$, and
find  $g_1$ and $g_2$. 

General solution of the equations of motion for the WZNW system can be
used to find the related solution for 2D induced gravity. 
To see that, we first rewrite Eqs.\eq{9} in the form
$\pd_{\pm} X_{\mp}=0$, where $X_+ =(G_+)_{12}/(G_+)_{22}$,
$X_-=-(G_-)_{21}/(G_-)_{22}$. Then, we use Eq.\eq{13} to obtain
\be
e^{\phi_1} ={4 \pd_+ X_+ \pd_- X_- \over\m (1- X_+ X_-)^2 }\, , \qquad
e^{\phi_2} ={4 \pd_+ Y_+ \pd_- Y_- \over\m (1- Y_+ Y_-)^2 }\, , \lab{15}
\ee
where $Y_+ =e_{11} X_+ +e_{12}$ and $Y_- =e_{11} X_- -e_{21}$. 
Finally, the solution for the induced gravity fields $\phi$ and 
$\bg_{\m\n}$ follows directly from Eq.\eq{12}. 

We expect that these results can be used to clarify our understanding
of topologically nontrivial solutions of the induced gravity \cite{6}.

\end{document}